\documentclass[12pt]{article}
\usepackage{float}
\usepackage{epsfig}
\usepackage{amsmath,amsthm,amssymb}
\usepackage{color}
\usepackage{textcomp}
\usepackage{slashed}
\usepackage{amsfonts}
\usepackage{dsfont}

\usepackage{calc}
\usepackage{ifthen}
\usepackage{graphics}
\usepackage{epsfig}
\usepackage{graphicx}
\usepackage{rotating}
\usepackage{subfigure}

\usepackage{multirow}
\usepackage{multicol}

\usepackage[colorlinks=false,linkcolor=blue]{hyperref} 
\hypersetup{colorlinks
,linkcolor=black
,filecolor=black
,urlcolor=black
,citecolor=black}

\usepackage[hang,small,bf]{caption}
\newcommand{\tr}{\mathrm{Tr~}}
\newcommand{\re}{\mathrm{Re }}
\newcommand{\pbp}{\ensuremath{\langle \bar \psi \psi \rangle}}
\newcommand{\asq}{\ensuremath{\langle A^2 \rangle}}
\newcommand{\msbar}{\ensuremath{\overline{\mathrm{MS}}}}
\newcommand{\lms}{\ensuremath{\Lambda_{\overline{\mathrm{MS}}}}}

\makeindex

\begin{document}

\vspace{-0.4cm}
\hfill SFB/CPP-12-64, HU-EP-12/27, RM3-TH/12-14

\begin{center}
  \textbf{
\begin{Large} 
Quark mass and chiral condensate from the Wilson twisted
mass lattice quark propagator 
\end{Large}
  } 
\end{center}

\begin{center}
  \vspace{.1cm}
  Florian~Burger$^a$,  Vittorio~Lubicz$^{b,c}$,
  Michael~M{\"u}ller-Preussker$^a$, Silvano~Simula$^{b,c}$, Carsten~Urbach$^d$
\end{center}

\vspace{.8cm}
\begin{center}
  \includegraphics[draft=false]{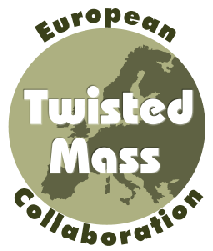}
\end{center}
\vspace{0.8cm}
\noindent
$^a$\,Humboldt-Universit\"at zu Berlin, Institut f\"ur Physik, 12489~Berlin, Germany\\
$^b$\,Dip. di Fisica, Universita Roma Tre, Via della Vasca Navale 84,
I-00146 Roma, Italy\\
$^c$\,INFN, Sezione di Roma Tre, Via della Vasca Navale 84, I-00146 Roma, Italy\\
$^d$\,Universit\"at Bonn, HISKP and Bethe Center for Theoretical
Physics,\\ $~~~$Nussallee 14-16, 53115~Bonn, Germany

\vspace{.4cm}
\centerline{October 2, 2012}

\begin{abstract}
In this work, we report about the determination of nonperturbative
OPE parameters from fits of continuum perturbation theory to the
Landau gauge quark propagator. The propagators are computed
numerically using lattice QCD with $N_f=2$ dynamical Wilson twisted
mass fermions. We use four different values of the lattice spacing ranging from 
$a \approx 0.1 \mathrm{~fm}$ to $a \approx 0.05 \mathrm{~fm}$ as well as several 
quark masses per lattice spacing. This allows us to obtain continuum results
for the chiral condensate and the up/down quark mass extrapolated to the
physical point. The main results are the average up/down quark mass
$m_q^{\overline{\mathrm{MS}}}(2 \mathrm{~GeV}) = 3.0 ~ (4) ~ (2) \
\mathrm{~MeV}$ at the physical point and 
$\langle\bar\psi\psi\rangle^{\overline{\mathrm{MS}}}(2 \mathrm{~GeV}) =
  -(299 ~ (26) ~ (29) \mathrm{~MeV})^3$ in the chiral limit.
We have also studied nonperturbative contaminations of our results at 
small values of the momenta, which are often interpreted as the contribution 
of the gluon condensate $\langle A^2 \rangle$. We do see contributions 
from such terms, which are, however, not stable over the  order in 
perturbation theory. 

\vspace*{0.5cm}\noindent
PACS numbers: 11.15.Ha, 12.38.Ge, 12.38.Aw \\
Keywords: Lattice QCD, quark propagator, Landau gauge, condensates, quark mass
\end{abstract}

\section{Introduction}
\label{sec:intro}

Quantum Chromodynamics (QCD) describes the strong interaction part in
our current standard model of elementary particle physics. Due to 
its particular properties, predictions for fundamental parameters 
of QCD require nonperturbative methods. The main tool in this 
context is lattice QCD, which allows predictions from first principles. 

Also perturbative calculations in QCD, where applicable, play
an important role for our understanding of QCD. For instance, in the
regime of large momenta two-point functions $\langle O(p)\
O(p')\rangle$ of some operator $O(p)$ can be written in terms of
an operator product expansion (OPE). The OPE contains 
coefficients carrying the dependence on the momenta, which can be
computed in perturbation theory as asymptotic series in powers of
the strong coupling $\alpha_s$, multiplied with local operators and 
appropriate factors of the quark mass. The matrix elements of those
local operators, like e.g.\ the chiral condensate, are of purely
nonperturbative nature.

A most natural combination of the two aforementioned notions is to
compute the momentum dependence of two point functions
nonperturbatively and compare these for large momenta with the OPE
prediction. Such a method not only allows us to compute 
estimates for fundamental parameters of QCD, such as quark masses and
condensates, but in principle also to determine the
strong coupling constant $\alpha_s$.

The quantity we consider in this work is the Landau gauge quark
propagator $P(k)$ in momentum space. Its OPE has the following
form
\begin{equation}
  \label{eq:sope}
  P(k) \sim \frac{1}{k^2} \left( \slashed{k} C_1(k^2) + C_m(k^2) m_q +
    C_{\bar\psi\psi}(k^2) \pbp + \ldots\right)\,.
  \end{equation}
The coefficient functions $~C_{X}(k^2),~~X\equiv 1, m, \bar\psi\psi,\dots~$
carry the whole momentum dependence
and can be computed in perturbative QCD for large $k^2$. The
nonperturbative information is encoded in the quark mass and the
condensate(s).

Since we are going to compute $P(k)$ using lattice QCD, we cannot
perform the calculation at arbitrary values for the momenta, since we
have to fulfil the inequality
\[
1/a^2 \quad \gtrapprox\quad  k^2 \quad \gtrapprox\quad  \Lambda_\mathrm{QCD}^2 \,.
\]
The first inequality ensures small lattice artifacts and the second one
the applicability of perturbation theory.
It is not a priori clear that such a window exists and one of
the questions we try to answer in this work is whether the values of
the lattice spacing available from state of the art lattice QCD
simulations are yet sufficient for such an investigation. Hence, we 
will pay special attention to both, lattice artifacts and nonperturbative
contaminations of our results.

Lattice artifacts have actually previously been a complication for
the applicability of the method we are going to apply: when Wilson
fermions are used naively in momentum space the leading contribution to
the OPE of the quark propagator is a constant term proportional to the
lattice spacing $a$. Even though this term will vanish eventually in
the continuum limit, at finite value of $a$ it will dominate the OPE,
since the mass and condensate contributions are suppressed by powers
of $1/k^2$. This makes the determination of the chiral
condensate difficult, whereas the quark mass can be
determined if several values of it are investigated~\cite{Becirevic:1999kb}. 
This complication can be circumvented by using the OPE of the pseudo-scalar 
vertex~\cite{Becirevic:2004qv} or by working in the 
$x$-space~\cite{Gimenez:2005nt}. 

Within the lattice formulation we are going to use here -- the so-called
Wilson twisted mass formulation of lattice QCD~\cite{Frezzotti:2000nk} --
such a term can be avoided due to automatic $\mathcal{O}(a)$
improvement~\cite{Frezzotti:2003ni}. In addition, we are going to
remove lattice artifacts of order $g_0^2a^2$ from our data as computed
in one-loop lattice perturbation theory~\cite{Constantinou:2009tr,Martha:2011}.

The procedure outlined above is not the only way to determine the
quark mass and the chiral condensate. They have been determined previously
for instance from fits of chiral perturbation theory to the data for
the pseudoscalar decay constant and mass, see for instance
\cite{Baron:2009wt}. But the analysis we are going to apply represents
an independent way to determine these important standard model
parameters with different systematics compared to other methods.

The main results of this paper are determinations of the 
average up/down quark mass at the physical point
and of the quark condensate, both within the
$\overline{\mathrm{MS}}$ scheme at renormalization scale $2 \mathrm{~GeV}$%
\footnote{The first errors are purely statistical and the 
second ones systematic, respectively.}
\begin{equation} 
\label{eq:final_mass}
  m_q^{\overline{\mathrm{MS}}}(2 \mathrm{~GeV}) = 3.0 ~ (4) ~ (2)  \mathrm{~MeV}
\end{equation}
and having performed the chiral limit  
\begin{equation} 
\label{eq:final_condensate}
  \langle\bar\psi\psi\rangle^{\overline{\mathrm{MS}}}(2 \mathrm{~GeV}) =
 -(299 ~ (26) ~ (29) \mathrm{~MeV})^3.
\end{equation}
For the chiral condensate we quote here the value stemming from one of
our fit strategies (fit $B$), as explained later. The other fit
strategies give slightly different, but compatible results.
We also discuss nonperturbative contaminations at small momenta, 
called two-dimensional gluon condensate $\langle A^2 \rangle$ and provide
effective values for it depending on the order of perturbation theory 
taken into account in the fits. 

The investigation presented in this paper is based on gauge
configurations as produced by the European Twisted Mass collaboration
(ETMC) with $N_f=2$ quark flavors of Wilson twisted mass
fermions~\cite{Baron:2009wt,Boucaud:2007uk,Urbach:2007rt,Boucaud:2008xu}.
We refer the reader to these references for all the 
details of the simulations.

The paper is organized as follows. In Sec.~\ref{sec:pert} we
discuss the quark propagator in perturbation theory and in
Sec.~\ref{sec:lat} our lattice formulation. In
Sec.~\ref{sec:ana} we present our analysis strategy and in
Sec.~\ref{sec:res} the corresponding results. We conclude with a
summary. 

\section{The Quark Propagator in Perturbation Theory}
\label{sec:pert}

In perturbation theory the quark propagator
is known up to three loops \cite{Chetyrkin:betagamma}. 
Recently, the OPE of the renormalized momentum space quark propagator in 
Landau gauge has been performed in Ref.~\cite{Chetyrkin:2009kh} in 
the $\msbar$ scheme. The authors have included terms up to mass
dimension three in their calculations. According to its Lorentz
structure, $P(k)$ can be written as
\begin{equation}
  P(k) = \frac{1}{k^2} S(k^2) \mathds{1} 
  + \frac{\slashed{k}}{k^2} V(k^2)\, .
  \label{propope}
\end{equation}
Assuming that we can use the OPE, one gets the following expansion
for the scalar and vector form factors $S(k^2)$ and
$V(k^2)$~\cite{Chetyrkin:2009kh} (up to operators of dimension three): 
\begin{equation}
  \label{opescalar}
  \begin{split}
    \frac{1}{k^2} S(k^2) &= C_m(k^2) m_q + \frac{C_{m^3}(k^2)}{k^2} m_q^3 + \\
    &\hphantom{ = } + \frac{C_{m A^2}(k^2)}{k^2} m_q \asq
    + \frac{C_{\bar \psi \psi}(k^2)}{k^2} \pbp  \\ 
  \end{split}
\end{equation}
and
\begin{equation}
  \label{opevector}
  \frac{1}{k^2} V(k^2) = C_1(k^2) \mathds{1} + \frac{C_{m^2}(k^2)}{k^2} m_q^2 
  + \frac{C_{A^2}(k^2)}{k^2} \asq\ .
\end{equation}
While the quark mass $m_q$ and the chiral condensate
$\langle\bar\psi\psi\rangle$ are clearly of physical origin, the
existence of a gluon condensate $\asq$ is debatable. However, in the
OPE such a term is not excluded.
The Wilson coefficients $C_1$, $C_m$, $C_{A^2}$, $C_{m^2}$, $C_{\bar \psi \psi}$, 
$C_{m A^2}$ and $C_{m^3}$ are functions of the strong coupling
constant $\alpha_s(\mu)$. The expansion can be found in
Ref.~\cite{Chetyrkin:2009kh}. The value of $\alpha_s(\mu)$ can be
computed via renormalization group (RG) evolution as described in
Ref.~\cite{Chetyrkin:rundec}, which needs $\lms$ as an input. For this
purpose we use a literature value, as
discussed later on.

Although known up to a certain order in $\alpha_s$, the perturbative
series of the Wilson coefficients in Eqs.~(\ref{opescalar}) and
(\ref{opevector})  may be truncated further in order to  
study the systematic effect of the truncation. For this purpose we 
stop the evaluation of the perturbative series at the $n_\mathrm{max}$'th power in 
$\alpha_s$ including at maximum terms of order $\alpha_s^{n_\mathrm{max}}$:
\begin{equation}
 C_{X} ~=~ C^{0}_{X} ~ + ~ C^{1}_{X}\alpha_s^{1} ~+~ \ldots 
       ~+~ C^{n_\mathrm{max}}_{X} \alpha_s^{n_\mathrm{max}}\,, 
\end{equation} 
where $X$ stands for $1,\, A^2,\, m^2 , \ldots$.
The truncation is consistently done in the Wilson coefficients as well 
as in the evolution of $\alpha_s$ by means of the $\beta$-function.

\section{Lattice Formulation}
\label{sec:lat}

The lattice quark propagator is calculated within the twisted mass
formulation of $N_f=2$ QCD
\cite{Baron:2009wt,Boucaud:2007uk,Urbach:2007rt,Boucaud:2008xu}. 
For a review see Ref.~\cite{Shindler:2007vp}. The
gauge action used to generate the ensembles was the tree 
level Symanzik improved gauge action (tlSym)~\cite{Weisz:1982zw}, viz.
\[
S_g = \frac{\beta}{3}\sum_x\left(  b_0\sum_{\substack{
    \mu,\nu=1\\1\leq\mu<\nu}}^4\{1-\re\tr(U^{1\times1}_{x,\mu,\nu})\}\Bigr. 
\Bigl.\ +\ 
b_1\sum_{\substack{\mu,\nu=1\\\mu\neq\nu}}^4\{1
-\re\tr(U^{1\times2}_{x,\mu,\nu})\}\right)\,,
\]
with the bare inverse gauge coupling $\beta=6/g_0^2$, $b_1=-1/12$ and
$b_0=1-8b_1$. The fermion action in the so-called twisted basis is
given by:
\begin{equation}
  S_F = a^4 \sum_x \bar \chi_x \left ( D_{\mathrm{W}} + m_0 + i \mu_q
    \gamma_5 \tau_3 \right ) \chi_x \equiv a^4  
  \sum_x \bar \chi_x D_{\mathrm{tm}} \chi_x\,.
\end{equation}
Here $D_{\mathrm{W}}$ represents the lattice Wilson Dirac operator, $m_0$ is
the usual bare quark mass and $\mu_q$ is the bare twisted quark mass, 
which is multiplied by the third Pauli matrix $\tau_3$ acting in flavor
space. Twisted mass fermions are said to be at \emph{maximal
twist} if the bare untwisted quark mass $m_0$ is tuned to its critical
value $m_\mathrm{crit}$, the situation we are working in. At
maximal twist, the twisted quark mass $\mu_q$ is related directly to
the physical quark mass and renormalizes multiplicatively only. Many
mixings under renormalization are expected to be simplified
\cite{Frezzotti:2003ni,Frezzotti:2004wz}. And -- most
importantly -- as was first shown in Ref.~\cite{Frezzotti:2003ni},
physical observables are automatically $\mathcal{O}(a)$ improved
without the need to determine any operator-specific improvement
coefficients. For details on tuning to maximal twist we refer the
reader to Ref.~\cite{Baron:2009wt}. The aforementioned twisted basis
$\bar\chi, \chi$ is at maximal twist related to the standard physical
basis $\bar\psi,\psi$ via the axial chiral rotation
\begin{equation}
  \label{eq:rotation}
  \psi = e^{i\pi\gamma_5\tau_3/4}\ \chi\, ,\qquad \bar\psi =
  \bar\chi\ e^{i\pi\gamma_5\tau_3/4}\,.
\end{equation}
In this framework we compute the Landau gauge twisted quark propagator
$P_\mathrm{tm}(x)$ in position space
\begin{equation}
  \label{quarkpropagator_tm}
  P_{\mathrm{tm}}(x) = \langle \chi_x \bar \chi_0 \rangle 
  = \langle (D_{\mathrm{tm}})^{-1} \rangle_{U}\,,
\end{equation}
where $\langle \ldots \rangle_{U}$ denotes the average over gauge 
field configurations $\{U\}$ 
collected by the ETMC, which were gauge fixed
to Landau gauge using the overrelaxation method described in
\cite{mandulaogilvie}.%
\footnote{Gribov copy effects have been found to be small for large 
momenta in studies of the lattice gluon and ghost propagators in
Ref.~\cite{Sternbeck:2005tk}. We have also carried out a study on the
Gribov copy dependence of the quark propagator on a test ensemble with
smaller lattice size and have found no ambiguities there.} 
The quark propagators calculated on these configurations are Fourier
transformed to momentum space and then
rotated into the physical basis using Eq.~(\ref{eq:rotation}) yielding
\begin{equation}
  P(k) =  \frac{1}{\sqrt{2}} \left (\mathds{1} + i \gamma_5 \tau^3
  \right )  
  P_{\mathrm{tm}}(k) \frac{1}{\sqrt{2}} \left (\mathds{1} + i \gamma_5
    \tau^3 \right )\,. 
\end{equation}
The details of the ETMC ensembles we used can be found in
Table~\ref{tab_latticesettings}. In total we consider four values of
the inverse gauge coupling $\beta$ corresponding to values of the
lattice spacing ranging from $0.1\ \mathrm{fm}$ to about $0.051\
\mathrm{fm}$~\cite{Baron:2009wt}, with a statistics of $240$ gauge
configurations for most of the ensembles considered. The values of the
(charged) pseudoscalar mass range from $600 \mathrm{~MeV}$ down to $250
\mathrm{~MeV}$. For setting the scale we use the results published by
ETMC in Ref.~\cite{Baron:2009wt}. 

\begin{table}[t]
  \addtolength{\tabcolsep}{-2pt}
  \centering
  \begin{tabular*}{.8\textwidth}{@{\extracolsep{\fill}}lccccc}
    Ensemble & $\beta$ &$L^3 \times T$  & $a \mu_q$      & $a$\ [fm] & $N_{\mathrm{conf}}$\\
    \hline \hline
    $A_1$ & $ 3.80 $ & $24^3\times 48$  & $0.0060$    & $\approx 0.10 $  & $231$ \\
    $A_2$ & $      $ &                  & $0.0080$    &                  & $240$ \\
    $A_3$ & $      $ &                  & $0.0110$    &                  & $240$ \\
    $A_4$ & $      $ &                  & $0.0165$    &                  & $240$ \\
    \hline
    $B_1$ & $ 3.90 $ & $24^3\times 48$  & $0.0040$    & $\approx 0.085$  & $240$ \\  
    $B_2$ & $      $ &                  & $0.0064$    &                  & $240$ \\ 
    $B_3$ & $      $ &                  & $0.0085$    &                  & $240$ \\
    $B_4$ & $      $ &                  & $0.0100$    &                  & $240$ \\
    $B_5$ & $      $ &                  & $0.0150$    &                  & $240$ \\
    \hline 
    $C_1$ & $ 4.05 $ & $32^3\times 64$  & $0.0030$    & $\approx 0.067$   & $240$ \\ 
    $C_2$ & $      $ &                  & $0.0060$   &                    & $161$ \\
    $C_3$ & $      $ &                  & $0.0080$   &                    & $163$ \\
    \hline
    $D_1$ & $ 4.20 $ & $48^3\times 96$  & $0.0020$    & $\approx 0.051$   & $188$\\
    $D_2$ & & $32^3\times 64$  & $0.0065$   &                   & $200$\\
    \hline
  \end{tabular*}
  \caption{Details of the ETMC gauge ensembles, used in the
    analysis. See Ref.~\cite{Baron:2009wt} for more details.}
  \label{tab_latticesettings}
\end{table}

On the lattice, Eq.~(\ref{propope}) is valid only up to lattice
artifacts. In particular, since parity is not a good symmetry of
Wilson twisted mass fermions at finite values of the lattice
spacing, a parity violating term in the propagator is allowed,
which is formally $\mathcal{O}(a)$. Ignoring higher order lattice
artifacts including $O(4)$ symmetry breaking terms, the lattice quark
propagator can be written as follows~\cite{Constantinou:2010gr}
\begin{equation}
  \label{formfactors}
  P(k) = - i \frac{\slashed{k}}{k^2} V(k^2) + \frac{1}{k^2}S(k^2) 
  + i \gamma_5 \tau^3\frac{1}{k^2}G(k^2)\,. 
\end{equation}
The parity violating term $G(k^2)$ comes with opposite sign for up and
down propagators. Averaging over up and down propagators hence
eliminates these artifacts~\cite{Constantinou:2010gr}. 
Note that the parity violating $\mathcal{O}(a)$ term encoded in $G(k^2)$
contributes to $S(k^2)$ for standard Wilson fermions.

Rotational symmetry is broken at finite values of the lattice
spacing. In order to reduce the impact of those artifacts on our
results, we carried out the following steps: firstly, we use for the
analysis the lattice tree level momenta of the quark propagator, 
$k_{\mu} = \frac{1}{a} \sin(a \tilde k_{\mu} + \frac{1}{2} \delta_{\mu
  0}  )$ with $\tilde k_{\mu} = (2 \pi n_{\mu})/(a L_{\mu})$.
Secondly, we restrict ourselves to momenta lying near the lattice
diagonal (i.\ e.\ off-axis) by applying 
the so-called cylinder cut \cite{Leinweber:1998uu}
\begin{equation}
  \label{eq:cylindercut}
  \begin{split}
    \left (  \sum_{\mu} \left ( \frac{n_\mu}{L_\mu} \right )^2 \right )
    -  
    \left (  \sum_{\mu} (\frac{n_\mu \hat N_\mu }{L_\mu}) \right )^2
    \le \frac{C_{\mathrm{cyl}}}{L^2}\,, 
  \end{split}
\end{equation}
where $\hat N_\mu = 0.5\cdot (1,1,1,1)$ is the lattice diagonal. The constant
has been set to $C_{\mathrm{cyl}}=1.6$ for the
large lattice at $\beta = 4.2$ and to $C_{\mathrm{cyl}}=1.1$ elsewhere. 
Following this procedure, the data for $S(k^2)$ and $V(k^2)$ show a
sufficiently smooth behavior.

Finally, we correct the form factors for lattice artifacts of
order $\mathcal{O}(g_0^2 a^2)$. These corrections have been computed
explicitly in Ref.~\cite{Constantinou:2009tr,Martha:2011} within one-loop 
lattice perturbation theory.

\section{Analysis}
\label{sec:ana}

\begin{figure}[t] 
  \centering
  \subfigure[]
  {\label{fig:1a}\includegraphics[scale=1]{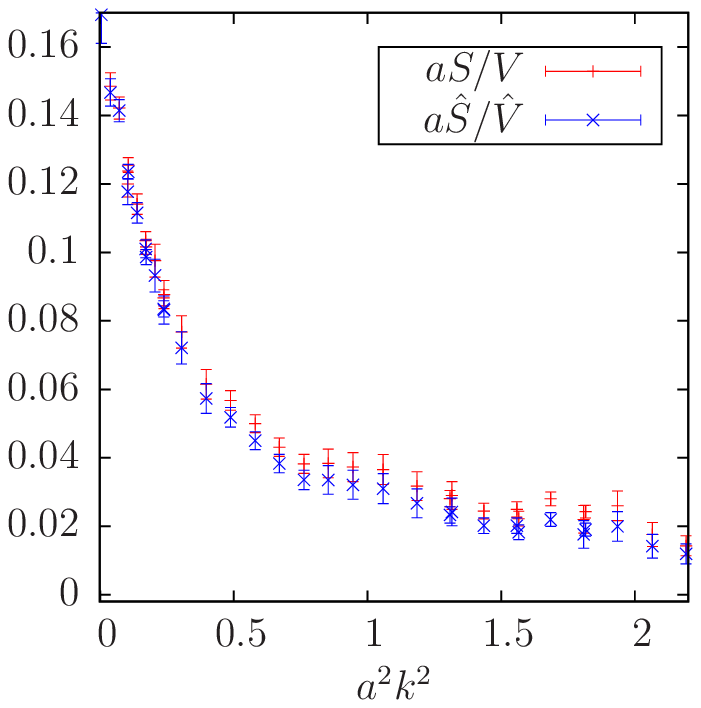}}\quad%
  \subfigure[]
  {\label{fig:1b}\includegraphics[scale=1]{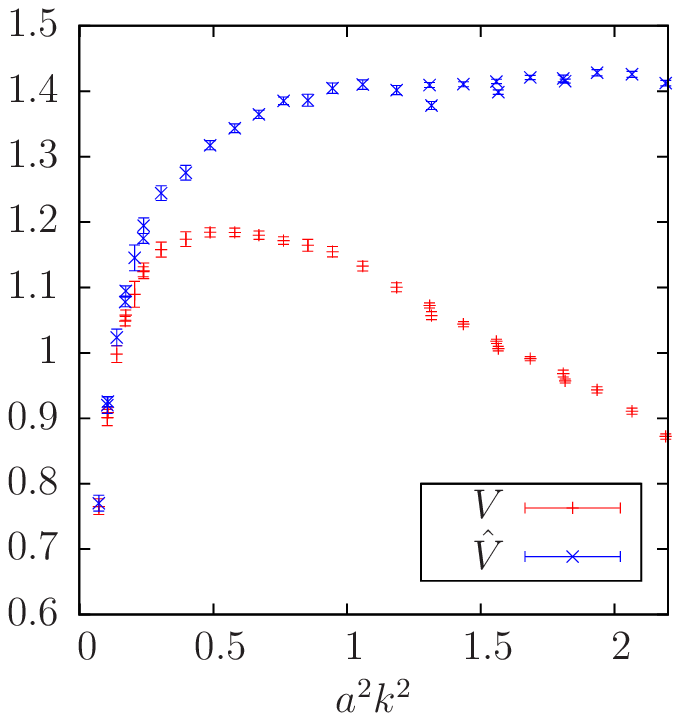}}
  \caption{Effect of the correction for $\mathcal{O}(g_0^2 a^2)$ 
    artifacts on $S/V$ (left) and on $V$ (right) for $\beta=3.8$. The
    effect becomes weaker for larger $\beta$ as expected.}
  \label{fig:1}
\end{figure}

\begin{figure}[t] 
  \centering
  \subfigure[]
  {\label{fig:2a}\includegraphics[scale=1]{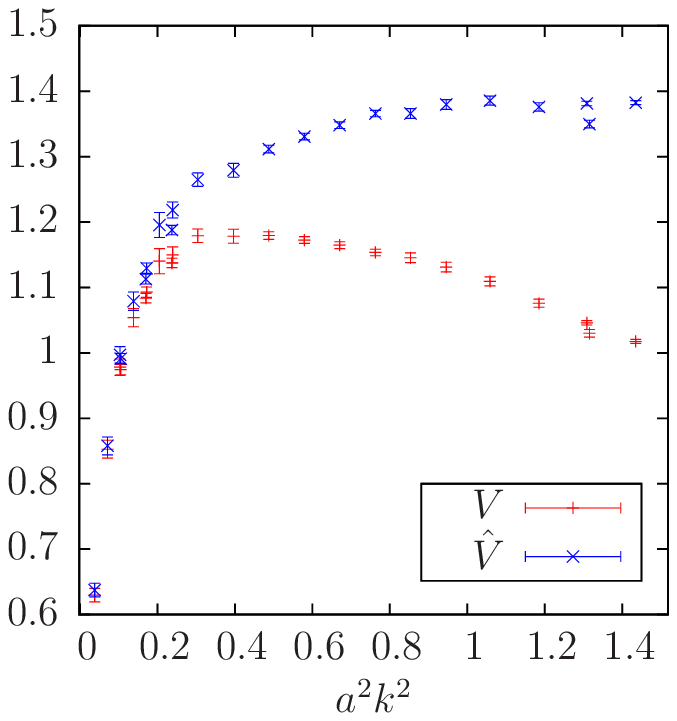}}\quad%
  \subfigure[]
  {\label{fig:2b}\includegraphics[scale=1]{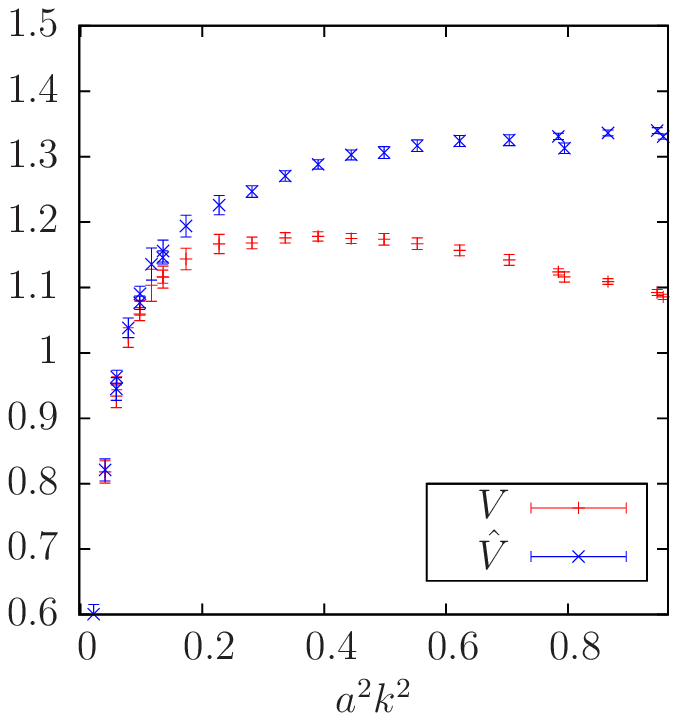}}
  \caption{{Same as in Fig.~\ref{fig:1b} but for $\beta=3.9$ (left) 
  and $\beta=4.05$ (right). The $x$-axis always displays the same 
  physical momentum range such that these figures may be compared.}}
  \label{fig:2}
\end{figure}

The basic quantities considered in our analysis are the bare form
factors $V(k^2)$ and $S(k^2)$ of the lattice quark propagator
Eq.~(\ref{formfactors}) after applying the cylinder cut and after
removing lattice artifacts of the order $\mathcal{O}(g_0^2 a^2)$. We
shall denote these \emph{corrected} form factors with $\hat S$ and
$\hat V$. The $\mathcal{O}(g_0^2 a^2)$ artifacts have
only modest effect on the ratio $S/V$ (after performing the cylinder
cut) while $V$ alone receives substantial corrections. As an example 
we show in Fig.~\ref{fig:1} uncorrected and corrected 
data for $aS/V$ and $V$ for $\beta=3.8$ in the left and right panel,
respectively. At this (smallest) $\beta$-value one sees dramatic
corrections to $V$. However, as visible in
Fig.~\ref{fig:2}, the corrections decrease as the
continuum limit is approached.

Our goal is to determine the renormalized quark mass and the chiral
condensate. Looking at the OPE's Eqs.~(\ref{opescalar}) and
(\ref{opevector}) it seems appropriate to study $\hat S(k^2)$, since
the quark mass can be determined from its leading quark mass
dependence and the chiral condensate is the leading contribution in the
chiral limit. However, if one wants to avoid the usage of
renormalization constants, it appears to be useful to consider the
scalar-to-vector form factor ratio
\begin{equation}
  \label{eq:ratio}
  \hat R(k^2) \equiv \frac{ \hat S(k^2)}{\hat V(k^2)}\,,
\end{equation}
since the renormalization constant cancels out. $\hat S$ and $\hat V$
are then replaced by their perturbative series and the resulting
expression is fitted to our numerical data for $\hat R(k^2)$ with the
renormalized quark mass, the renormalized chiral condensate and
possibly further terms as fit parameters. We remark in passing that as
soon as the renormalization constants in $\hat R$ cancel in between
nominator and denominator, the OPE of $\hat R$ is written in renormalized
quantities only. Scheme and renormalization scale depend then on the
scheme and scale the perturbative expansion is performed in.

For the fits of the ratio $\hat R$ as well as of the formfactors 
$\hat S$ and $\hat V$ separately we have followed three
different strategies.

\newpage
\begin{enumerate}
\item Fit $A$:\\
  The fits of Eq.~(\ref{eq:ratio}) to the data are performed
  simultaneously to all ensembles at fixed $\beta$-values,
  allowing for a different fit parameter value $m_q$ for each
  ensemble, but only for one global fit parameter corresponding to
  $\langle\bar\psi\psi\rangle$. All other contributions from the OPE
  are neglected. At each value of 
  $\beta$ the results for $m_q$ are then interpolated to reference 
  values of $r_0 m_\mathrm{PS}$, 
  $r_0$ denoting the so-called Sommer scale \cite{Sommer:scale} 
  and $m_\mathrm{PS}$ the pseudo-scalar meson mass, respectively. 
  Those interpolated results are extrapolated to the
  continuum limit and then to the physical point where appropriate.

\item Fit $B$:\\
  Resigning to determine a value for the renormalized quark mass, we
  extrapolate our data for $\hat R(k^2)$ to the chiral limit at each
  $\beta$-value first. Only then we fit the perturbative series to the
  data at each $\beta$-value, with terms proportional to powers of
  $m_q$ set to zero and still neglecting all other contributions from
  the OPE. Thereafter the result for $\langle\bar\psi\psi\rangle$ is
  extrapolated to the continuum limit.

\item Fits $A'$ and $B'$:\\
  At small values of $k^2$ our data is potentially contaminated by
  additional nonperturbative terms in the OPE. One example is what is
  often called the gluon condensate $\langle A^2\rangle$. We shall avoid
  here the discussion of its mere existence (see also 
  \cite{Boucaud:2005rm,Blossier:2010ky,Pene:2011kg}), but we shall investigate
  whether our data is contaminated by effects that may effectively
  look like the gluon condensate. Since including such a term 
  into the fits of type $A$ or $B$ applied to the ratio $\hat R$
  appears to be not stable, we first determine $\langle A^2\rangle$
  from $\hat V(k^2)$ only, and use it afterwards as an input for a fit of
  type $A$ and $B$.

\item Fit $C$:\\
  As a further check of the sensitivity of the extraction of the quark and 
  gluon condensates against different assumptions in the fitting procedures, 
  we resign to determine the quark masses $m_q$ (as in fits $B$ and $B'$) 
  and instead we fix their values to the ones of the renormalized quark 
  masses, which at maximal twist  are given by the twisted bare quark 
  masses, $a \mu_q$, multiplied by $1 / Z_P$, where $Z_P$ is the 
  renormalization constant of the pseudoscalar current\footnote{The values 
  of $Z_P$ in the $\msbar$ scheme at a renormalization scale of 
  $2 \mathrm{~GeV}$ can be read off from Ref.~\cite{Constantinou:2010gr} 
  at $\beta = 3.80, 3.90, 4.05$ and from Ref.~\cite{Blossier:2010cr} at 
  $\beta = 4.20$.}. In what follows, we will refer to such values as 
  the $Z_P$-based quark masses.
  
  Contrary to the previous fits $A$ and $B$, we do not use the data for 
  the ratio given by Eq.~(\ref{eq:ratio}), but instead the 
  data for the two form factors, $\hat S$ and $\hat V$, separately in order to 
  improve the sensitivity to the value of the gluon condensate, and 
  moreover we consider simultaneously all the data for the four values 
  of the lattice spacing. To this end we have to consider explicitly 
  the values of the renormalization constant of the quark field, $Z_q$, 
  which will be treated as free parameters, and we introduce also simple 
  discretization terms, proportional to the square of the lattice spacing, 
  for the quark masses and the condensates (see next section).   
\end{enumerate}

We perform fully correlated fits using the inverse covariance matrix
in our $\chi^2$-functions as described in
Ref.~\cite{Michael:correlated}. In order to estimate the systematic
error induced by a specific choice of the fit range we have
performed all fits in several fit ranges. The results we
give are then weighted according to the $\chi^2$-distribution function
and our final result consists of the weighted average over all fit
ranges~\cite{Baron:2009wt}. The fit ranges are consistently chosen
among different values of the lattice spacings in such a way that the
physical momentum range in terms of $r_0^2 k^2$ is kept approximately
constant. The different sets of fit ranges we used are
\begin{equation}
  \begin{split}
    & \ \ \ \ \ \ \ \ \ \ \ \ r_0^2 k^2  \in [18, \ldots, 58], 
                        [18, \ldots, 64], [18, \ldots, 66], \\
    & \hphantom{\ \ \ \ \ \ \ \ \ \ \ \ \ r_0^2 k^2  \in} [19, \ldots, 64], 
                        [20, \ldots, 64], [21, \ldots, 64], \\ 
  \end{split}
  \label{eq_fit_A_fitrange2}
\end{equation}
which in physical units lie within the range
\begin{equation}
  3.9 \mathrm{~GeV}^2 \le k^2 \le 14.6 \mathrm{~GeV}^2\,.
  \label{eq_fit_A_fitrange2physical}
\end{equation}
The statistical errors are estimated using a bootstrap procedure to  
propagate the errors consistently to the next step of the analysis. 
In order to estimate the systematic error related to the truncation in 
the perturbative series we carried out the fits using $n_\mathrm{max}=2$ and
$n_\mathrm{max}=3$ and we take the difference in the results as systematic
uncertainty. 

The evaluation of the perturbative series requires a value of $\alpha_s$
as input, for which a value of $\lms$ is needed. For this purpose we use
the value $\lms=0.330 ~ (23) \mathrm{~GeV}$~\cite{Blossier:2010ky} (see
also Table~\ref{tab:fitinput})
which is in good agreement with other $N_f=2$ determinations of the 
same quantity~\cite{Jansen:2011vv,Fritzsch:2012wq}.
The error of this number is taken into account in our
bootstrap analysis and contributes to the statistical errors of our fit results.

\begin{table}
  \centering
  \begin{tabular}[htp]{c c c }
    $m_{\pi^0}$ [MeV] & $r_0$ [fm]  & $\lms$ [GeV] \\
    \hline
    \hline 
    134.9766(6) & 0.42(2)  &  0.330(23) \\
  \end{tabular}
  \caption[Input parameters for the fits]{Physical quantities 
    used as an input parameters for the fits and analysis. The values
    of $m_{\pi^0}$ and $r_0$  have been taken from
    Refs.~\cite{Yao:2006px,Baron:2009wt}. The value of $\lms$ is
    taken from Ref.~\cite{Blossier:2010ky}.} 
  \label{tab:fitinput}
\end{table}

\section{Results}
\label{sec:res}


\subsection{Fit $A$: Determination of Quark Condensate and Mass }
\label{sec:fit_A}

Ignoring terms proportional to $\langle A^2\rangle$ and higher order
terms in the OPE Eqs.~(\ref{opescalar})
and (\ref{opevector}), we have performed fits to the chirally  
non-extrapolated data of $\hat S/\hat V$ with the quark mass and the chiral
condensate as fit parameters only. The fits are performed as
discussed previously. Throughout this analysis the renormalization
scale has been fixed at $\mu = 2 \mathrm{~GeV}$ in the perturbative
series. 

For all fit ranges quoted above the fits have produced stable results
with  acceptable $\chi^2/\mathrm{dof}$ values ranging from $0.78$ to
$1.24$. We could not 
increase 
the lower boundary beyond $r_0^2 k^2 =
21$  as the fit for $\beta=3.8$ then turned unstable. 
In all considered fit ranges the fitted values of the quark mass 
and the chiral condensate have been compatible with each other 
within errors.

\begin{figure}[tb] 
  \centering
  \subfigure[]
  {\label{fig:3a}\includegraphics[scale=1]{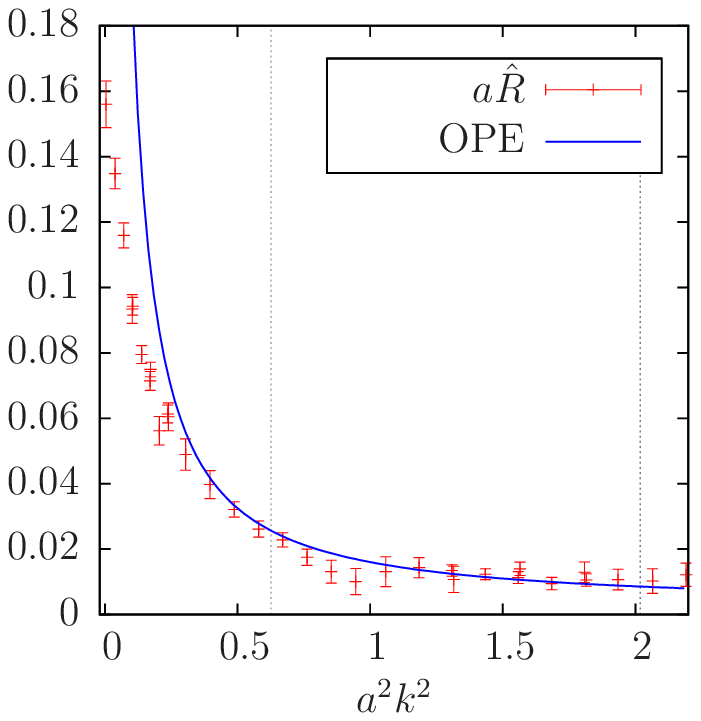}}\quad
  \subfigure[]
  {\label{fig:3b}\includegraphics[scale=1]{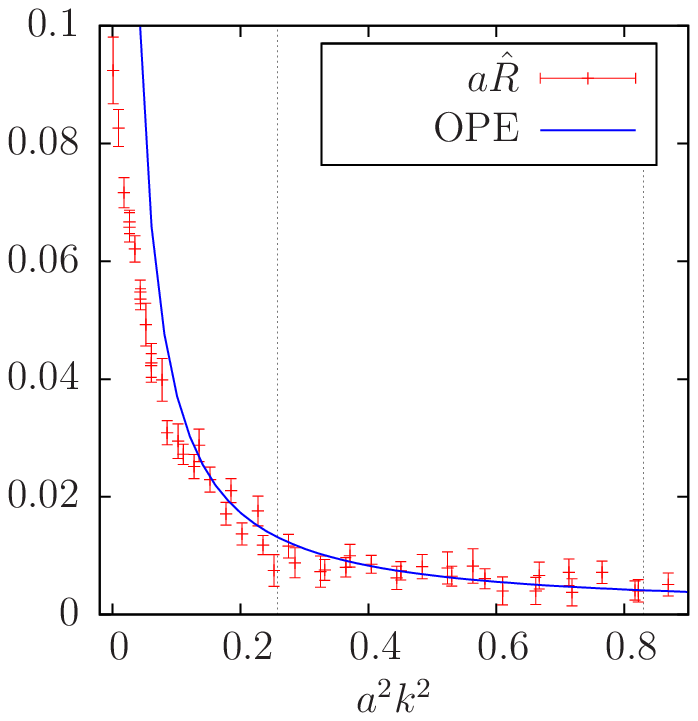}}
  \caption{Fits of $\hat R$ in lattice units to our data for ensemble
    $B_1$ at $\beta=3.9$ (left) and ensemble $D_1$ at $\beta=4.2$ (right).
    The vertical lines indicate the fit range. These plots
    correspond to fit strategy $A$.} 
  \label{fig:3}
\end{figure}

In Figs. \ref{fig:3a} and \ref{fig:3b} we show exemplary
fits for ensemble $B_1$ at $\beta=3.9$ and $D_1$ at $\beta=4.2$ with
dashed vertical lines indicating the chosen fit range.

\begin{figure}[tb] 
  \centering
  \subfigure[]
  {\label{fig:4a}\includegraphics[scale=1]{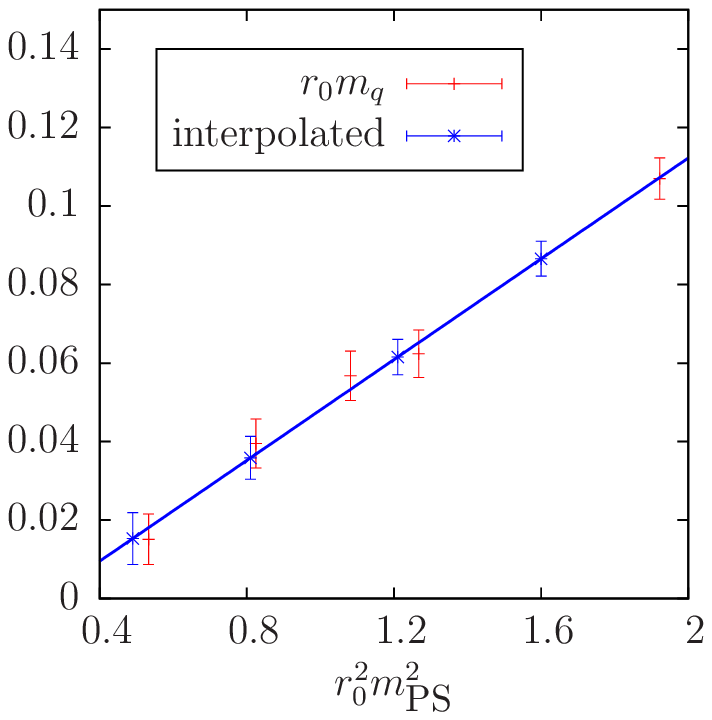} }\quad
  \subfigure[]
  {\label{fig:4b}\includegraphics[scale=1]{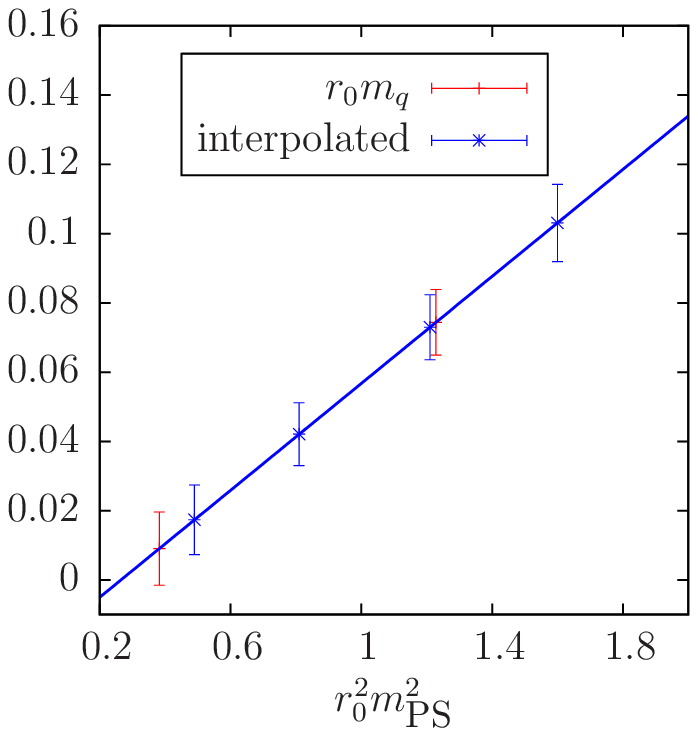}} \\
  \caption{Interpolation to the four chosen pion mass reference points 
    in $m_{\mathrm{PS}}^2$ for $\beta=3.9$ (left) and $\beta=4.2$ (right). 
    We also show the linear fit used for interpolation 
    to the reference points. The blue points correspond to the 
    interpolated values. These plots correspond to strategy fit $A$.
  }
  \label{fig:4}
\end{figure}

In order to perform a continuum extrapolation of the fitted 
quark masses $m_q$ we first have to interpolate (extrapolate)
the values to common reference points of the squared pseudoscalar mass
$m_{\mathrm{PS}}^2$ at each $\beta$-value. We have chosen the following
reference points 
\[
r_0^2 m_{\mathrm{PS}}^2 \in \{0.49,0.81,1.21,1.60\}\,,
\]
which allow us to use interpolations in the pseudoscalar mass in most of the
cases. Only for $\beta = 4.05$ and  $\beta = 4.2$ we have to
perform a short extrapolation for the largest reference mass  $r_0^2
m_{\mathrm{PS}}^2 = 1.60$ and for $\beta = 3.8$ and $\beta = 3.9$ we
have to rely on extrapolations for the smallest reference point.
In Fig.~\ref{fig:4} we show as an example the
interpolation  to these reference points for $\beta = 3.9$ and $\beta
= 4.2$. 

\begin{figure}[tb] 
  \begin{center}
    \subfigure[]
    {\label{fig:5a} \includegraphics[scale=1]{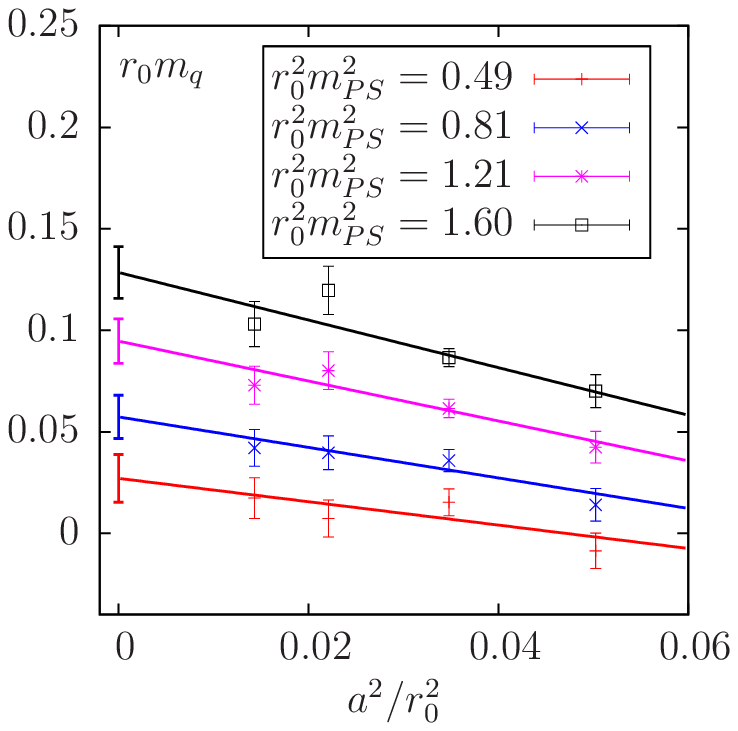} }
    \subfigure[]
    {\label{fig:5b} \includegraphics[scale=1]{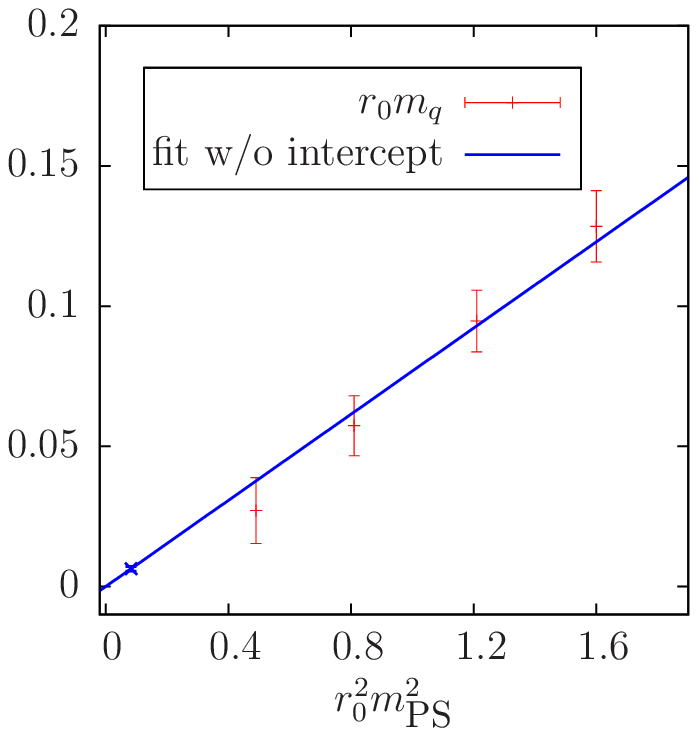} } \\
    \caption{Continuum limit of $r_0 m_q$  for the four chosen
      reference points (left) and extrapolation of the continuum 
      extrapolated quark mass values $r_0 m_q$ to the physical pseudoscalar
      mass (right). The linear fit has been constrained to 
      go through the origin. This result again corresponds to fit $A$.}
  \end{center}
  \label{fig:5}
\end{figure}

For each reference point we perform a separate 
continuum limit of the quantity $r_0 m_q$ in $a^2$ which 
is shown in Fig.~\ref{fig:5a}. 
The data appears to be compatible with a linear behavior in $a^2$ for
all chosen reference points as expected.

\begin{figure}[tb] 
  \begin{center}
     \includegraphics[scale=1]{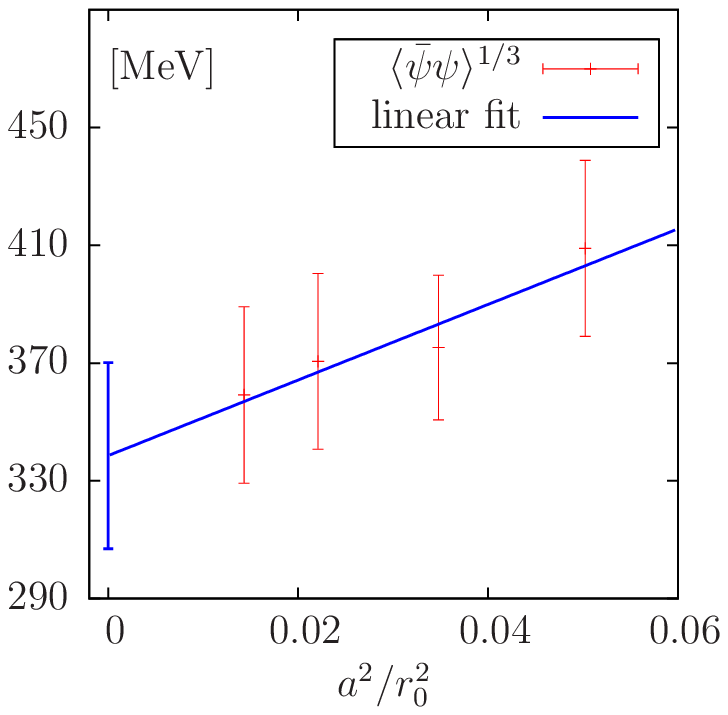}
  \end{center}
  \caption{Continuum limit of $\langle \bar \psi \psi \rangle$ from
    fit $A$.}
  \label{fig:6}
\end{figure}

Finally, the continuum quark mass data has to be extrapolated to the
physical pion mass $m_{\pi^0} = 134.9766 ~ (6)
\mathrm{~MeV}$~\cite{Yao:2006px}, for which we use a linear curve with
zero intercept (leaving the intercept free gives compatible results). 
The extrapolation is shown in Fig.~\ref{fig:5b}.
For $\langle \bar \psi \psi \rangle$ we have also performed a continuum 
extrapolation linear in $a^2$ as shown in
Fig.~\ref{fig:6}. Note that we also tried to
include chiral logs for the quark mass dependence of
$m_\mathrm{PS}^2$, however, at our current precision this does not
make a difference.
 
After performing a weighted average over the chosen fit ranges we
quote the following results for Fit A:
\begin{equation}
  \frac{\langle \bar \psi \psi \rangle^{\msbar}}{N_f} (2 \mathrm{~GeV})
  = -(335 ~ (37) ~ (35) \mathrm{~MeV})^3\, ,\quad  
  m_q^{\msbar}(2 \mathrm{~GeV}) = 3.0 ~ (4) ~ (2) \mathrm{~MeV}\,,
\end{equation}
where the first error is statistical and the second is a systematic 
error reflecting the uncertainties related to the fit range and to 
the truncation of the perturbative series. The systematic error due to the 
variation of the fit range is taken as the maximum deviation from the
$\chi^2$-averaged result. This amounts to about $0.08 \mathrm{~MeV}$
($(5 \mathrm{~MeV})^3$) systematic uncertainty for the quark mass
(condensate). Furthermore, decreasing the perturbative order to
$n_\mathrm{max}=2$ results in a smaller fitted quark mass value as well 
as a higher value for $\pbp$. The systematic errors we quote is the 
change of the central values when we apply this modification and added 
by the change caused by varying the fit range. 

\subsection{Fit $B$: Determination of $\pbp$ in the Chiral Limit}
\label{sec:fit_B}

\begin{figure}[tb] 
  \begin{center}
    \subfigure[]
    {\label{fig:7a}\includegraphics[scale=1]{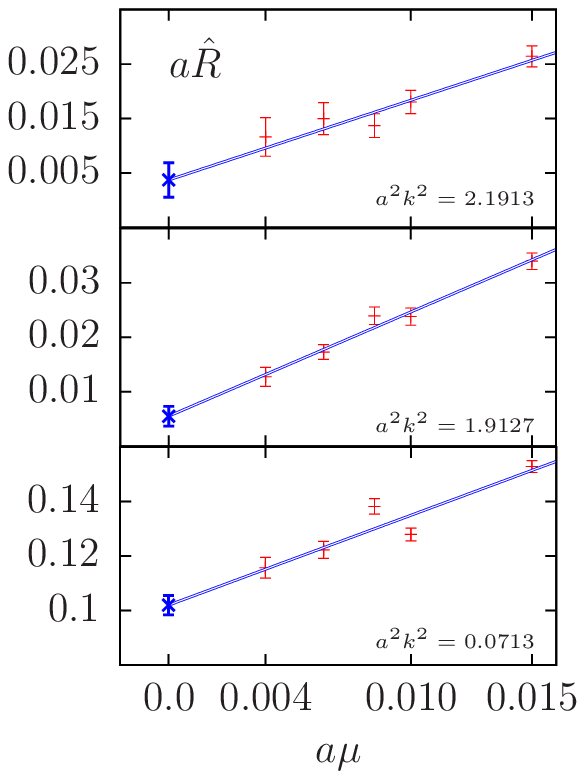} }
     \quad
    \subfigure[]
    {\label{fig:7b}\includegraphics[scale=1]{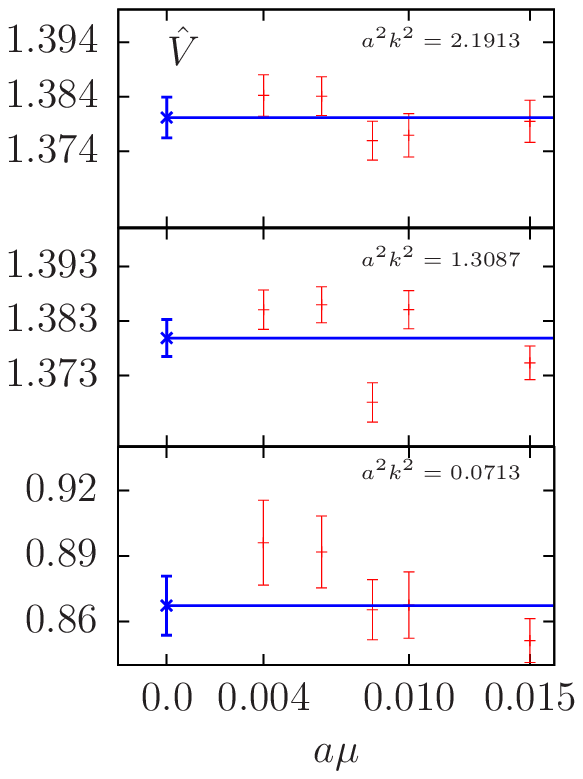} } \\ 
    \caption{(left) Chiral limit for $\hat R$ at
      $\beta=3.9$. Data has been extrapolated assuming a linear
      dependence on the bare quark mass $a \mu_q$ which is compatible
      with our data at every momentum considered. (right) Chiral limit
      for $\hat V$ at $\beta=3.9$ assuming a constant dependence on the
      bare quark mass $a \mu_q$. Here we follow fit strategy $B$. } 
  \end{center}
  \label{fig:7}
\end{figure}

As in the chiral limit the chiral condensate is the only
nonperturbative parameter (disregarding again a  
possible gluon condensate) we expect it to be estimated more reliably
and with less statistical error than in the finite mass case. 
Having calculated the quark propagator for two to five bare twisted
quark mass values $\mu_q$ we can 
perform a chiral extrapolation for each lattice spacing
separately. This limit has been performed linearly in in the bare
twisted quark mass $a \mu_q$ for $\hat R(k^2)$. Fig.~\ref{fig:7a} 
shows exemplary fits for $\beta = 3.9$ at three
representative values of $a^2 k^2$, one at the lower end, one in the
middle and one at the upper end of the considered momentum range. 
For any lattice spacing and any other momentum not shown here 
the data is consistent with such an extrapolation.   

\begin{figure}[tb] 
  \centering
    \subfigure[]
    {\label{fig:8a}\includegraphics[scale=1]{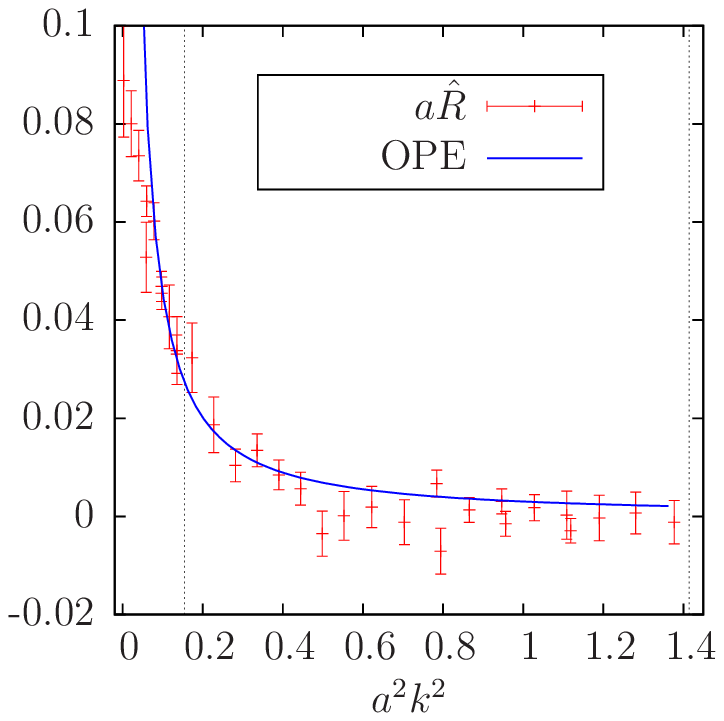} } 
    \subfigure[]
    {\label{fig:8b}\includegraphics[scale=1]{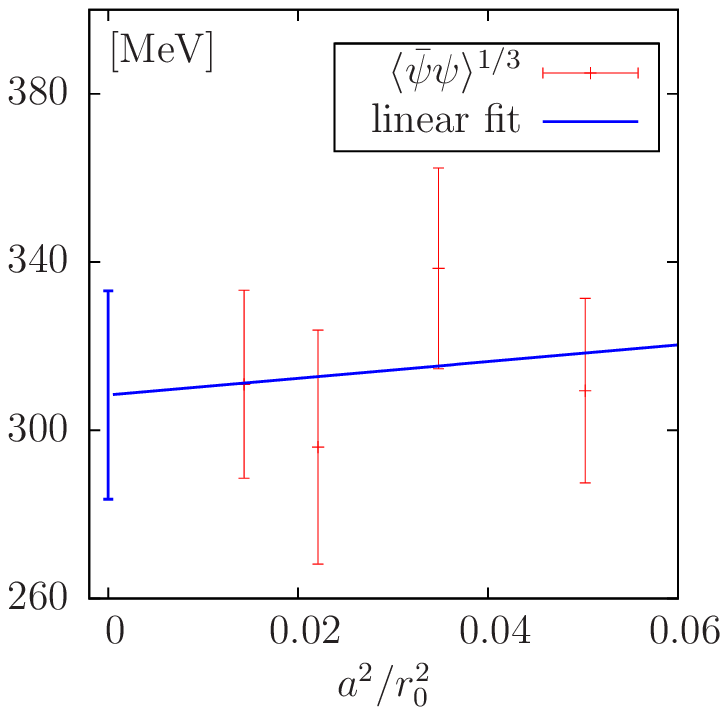}} \\ 
    \caption{(left) Fit of $a \hat R$ in the chiral limit using fit strategy
      $B$ for
      $\beta=4.05$. The vertical lines indicate the fit
      range. (right) Continuum limit extrapolation of $\pbp$. }
  \label{fig:8}
\end{figure}

We have then fitted the mass extrapolated data via
Eqs.~(\ref{opescalar}) and (\ref{opevector}) disregarding all other
OPE terms. The corresponding one parameter fits in $\langle \bar \psi \psi
\rangle$ have produced $\chi^2/\mathrm{dof}$ values in the range 0.8 to
1.8 and have been performed in the fit ranges: 
\begin{equation}
  \begin{split}
    & \ \ \ \ \ \ \ \ \ \ \ \ r_0^2 k^2  \in [7, \ldots, 64], 
                        [9, \ldots, 64], [12, \ldots, 64], \\
    & \hphantom{\ \ \ \ \ \ \ \ \ \ \ \ \ r_0^2 k^2  \in} [7, \ldots, 60], 
                        [7, \ldots, 58], [7, \ldots, 56].\\
  \end{split}
  \label{eq_fit_B_fitrange}
\end{equation}
Note that we had to extend the fit range compared to fit $A$ towards
the infrared in order to be sensitive to the curvature of the chirally
extrapolated data and to obtain stable results.
As an example we show the fit for $\beta=4.05$ in
Fig.~\ref{fig:8a}. The continuum extrapolation has again been
performed in the lattice spacing squared and is shown in 
Fig.~\ref{fig:8b}. After a weighted average over
the different fit ranges and the continuum extrapolation we get the
following result for the chiral condensate: 
\begin{equation}
   \frac{\langle \bar \psi \psi \rangle^{\msbar}}{N_f} (2 \mathrm{~GeV})
   =  - (299 ~ (26) ~ (29) \mathrm{~MeV})^3\,, 
   \label{eq_fit_B_result}
\end{equation}
where again the second error is systematic. Lowering the
perturbative order to $n_\mathrm{max} = 2$ results in a higher value of $\pbp$. 
The systematic error due to the use of different fit ranges is evaluated in the 
same manner as for fit $A$ and amounts to $(6 \mathrm{~MeV})^3$.

\subsection{Fits $A'$ and $B'$: Nonperturbative $A^2$ Contamination}
\label{sec:fit_AB_prime}

As discussed in the introduction, on the lattice we have to restrict
the analysis of the quark propagator to a window in the squared
momenta, as for too small momenta the perturbative expansion will not
be valid and for too large momenta lattice artifacts will be too
large. Of course, with any choice of this window one can never be sure
to be free of this sort of artifacts. For this reason we extend our 
analysis by including the further terms 
$m\langle A^2\rangle $ and $\langle A^2\rangle$ 
in the OPE. Unfortunately, it turns out that an inclusion of these terms as
fit parameters in the fit to our data for $\hat R(k^2)$ appears to be
not stable. This is why we follow a different strategy: we first
determine an estimate for $\asq$ from $\hat V(k^2)$ in the chiral limit
alone, and use this estimate to repeat the fits we discussed
before. This is not a fully consistent treatment, but it should
provide an estimate of the uncertainty in our results.

In more detail, we first studied $\hat V(k^2)$ of the quark propagator
according to Eq.~(\ref{opevector}) in the chiral limit. In this form
factor the dimension two term represents the first  
nonperturbative OPE contribution when the quark mass is extrapolated
to zero and thus one has a direct handle on this term.
A similar study has been done using the inverse lattice quark propagator
in Ref.~\cite{Blossier:2010vt}. Note that unlike $\hat R(k^2)$ the form
factor $\hat V$ requires renormalization, which we include as free
parameters into our fits.

\begin{figure}[htb] 
  \centering
  \subfigure[]
  {\label{fig:9a}
  \includegraphics[scale=1]{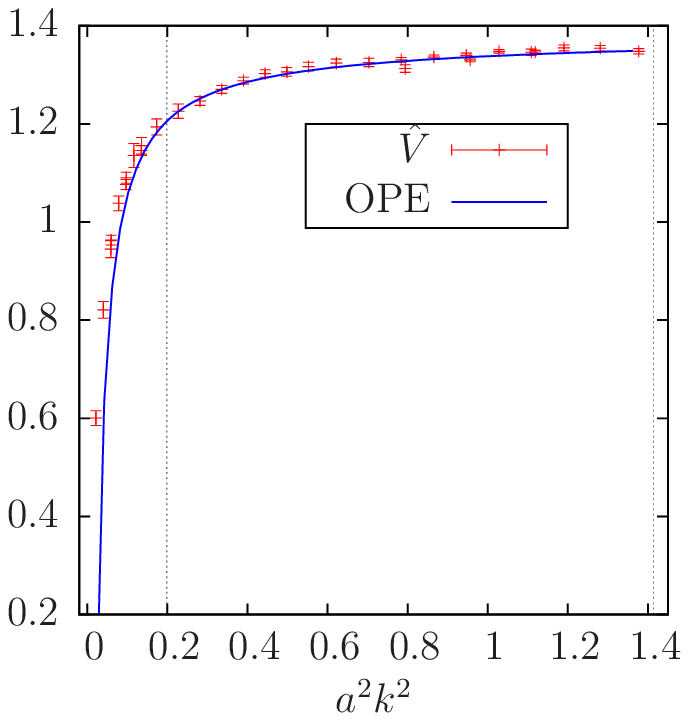} }\quad
  \subfigure[]
  {\label{fig:9b}
  \includegraphics[scale=1]{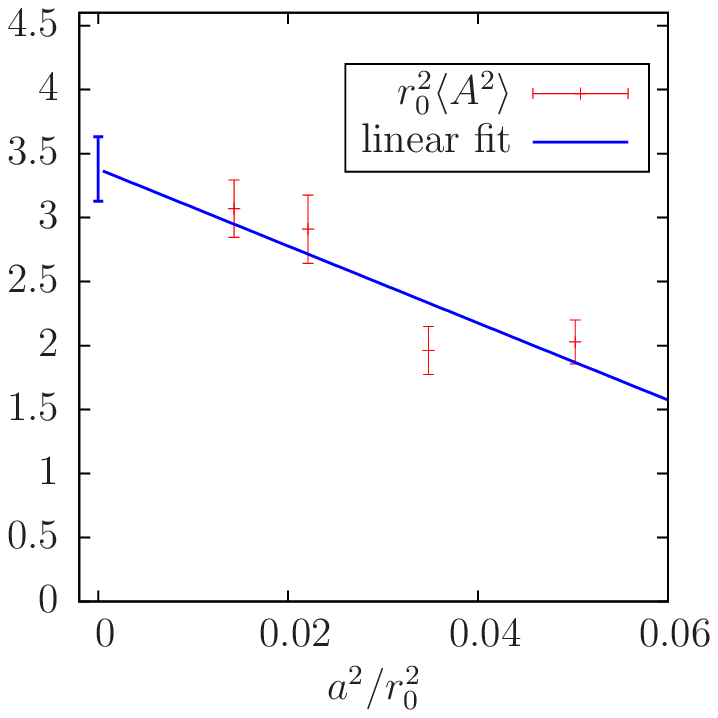} } \\
  \caption{(left) Chirally extrapolated vector form factor $\hat V$ and fits of
    perturbation theory to data at $\beta=4.05$.
    The vertical lines indicate the fit range. (right) The
    continuum extrapolation of $r_0^2 \asq$ for $n_\mathrm{max}=2$ 
    in $a^2/r_0^2$.} 
  \label{fig:9}
\end{figure}

The chiral limit of $\hat V(k^2)$ has been performed constant in $a
\mu_q$. This is justified as can be seen for instance in
Fig.~\ref{fig:7b} where we show the chiral limit of $\hat
V$ for $\beta=3.90$ for three values of $a^2k^2$. We then fit the OPE
formula, Eq.~(\ref{opevector}), to our data for $\hat V(k^2)$ in the 
same fit ranges as used for the fits $A$ and $B$. We obtain rather
large values for $\chi^2/\mathrm{dof}$ of about $\approx 5-7$ for
$\beta=3.8$ and $\beta = 3.9$ which are related to the fact that the
cut in momenta we have applied does not work well for two points of
$\hat V$ at $a^2 k^2 \approx 1.3$ and $a^2 k^2 \approx 1.55$ (see the
slight spread of the data at these 
regions in Fig.~\ref{fig:1b}). If we had discarded these points 
we would have obtained $\chi^2/\mathrm{dof} < 3.5$ in both cases.
The fits for the two smallest lattice spacings has yielded acceptable 
values of $\chi^2/\mathrm{dof}$ staying below $\approx 1.3 (1.5)$ for 
$\beta=4.05 (4.2)$ (see Fig.~\ref{fig:9a}).
Finally, the results have been extrapolated to the continuum limit
linearly in $a^2$, as shown in Fig.~\ref{fig:9b}.
For the gluon condensate in the continuum limit we obtain 
\begin{equation*}
      \asq^{\msbar}  = 0.65 ~ (09) ~ (17) \mathrm{~GeV}^2\,.
\end{equation*} 
The large systematic error is dominated by the truncation of the perturbative order.
In order to better understand the role of the $\asq$ term in the OPE,
we have studied the dependence of the values of $\asq$ on the
order of perturbation theory that has been used in
Eq.~(\ref{opevector}). To this end we have truncated perturbation
theory at the order $\alpha_s^{n_\mathrm{max}}$ and have performed fits with
$n_\mathrm{max}$ ranging from $1$ to $3$. For these fits we have restricted 
ourselves to only one fit range with $9 \le r_0^2 k^2 \le 64$. 
The resulting values of $\asq$ have then been extrapolated 
linearly in $a^2/r_0^2$ to the continuum, as shown in 
Fig.~\ref{fig:9b} for $n_\mathrm{max}=2$. 

\begin{figure}[htb]
  \begin{center}
    \includegraphics[scale=1]{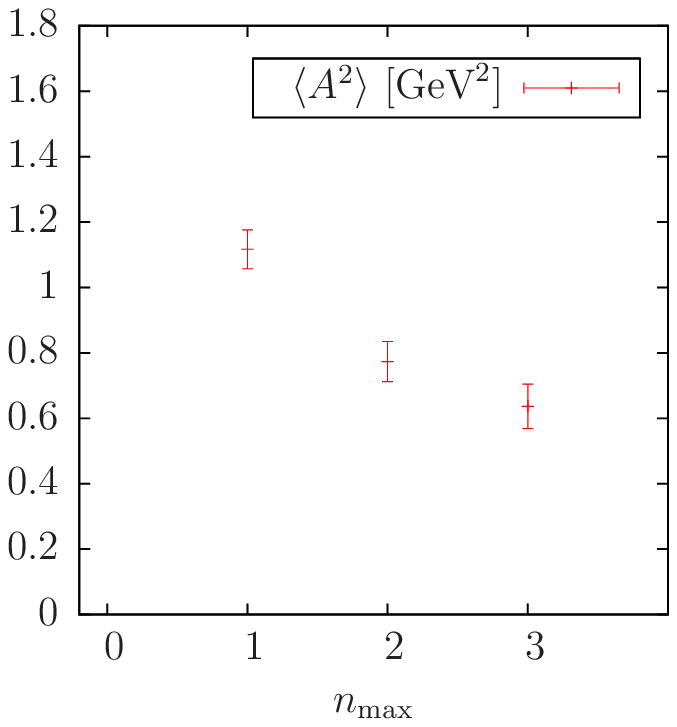} 
  \end{center}
  \caption{$n_\mathrm{max}$ dependence of the fitted gluon condensate $\asq$
    after continuum extrapolation.} 
  \label{fig:10}
\end{figure}

In Fig.~\ref{fig:10} we show the continuum extrapolated
expectation value of $\asq$ as a function of $n_\mathrm{max}$. From the figure
we conclude that with increasing $n_\mathrm{max}$ the continuum
value of $\asq$ decreases with no saturation visible (yet). Hence,
we might conclude that this dimension two term is effectively
describing higher order terms in perturbation theory which are not
included in our analysis, see Ref.~\cite{Narison:2009ag} for a
discussion. However, it seems that this contamination is not
negligible for our data when momenta as low as 
$r_0^2 k^2 ~\sim 9$ ($k^2 \sim 2 \mathrm{~GeV}^2$)
are included in the fit. 

Next we investigate the influence of this $\asq$ term on the results
for $\pbp$ and the quark mass. To this end we use the values of $\asq$
determined as discussed above as an input for fit with strategies $A$
and $B$  with fit ranges given in Eq.~(\ref{eq_fit_A_fitrange2}) and 
Eq.~(\ref{eq_fit_B_fitrange}), as explained previously.
The result of fit $B'$ is a slightly smaller value of $\pbp$ 
\begin{equation}
   \frac{\langle \bar \psi \psi \rangle^{\msbar}}{N_f} (2 \mathrm{~GeV})
   =  - (294 ~ (25) \mathrm{~MeV})^3\,, 
\end{equation}
which is however compatible within errors with the value from Fit $B$.
Also the results of fit $A'$ differ only slightly from fit $A$:
\begin{equation}
  \frac{\langle \bar \psi \psi \rangle^{\msbar}}{N_f} (2 \mathrm{~GeV})
  = -(324  ~ (37) \mathrm{~MeV})^3\,,
  \hspace{1 cm} m_q^{\msbar}(2 \mathrm{~GeV}) = 3.0 ~ (4) \mathrm{~MeV}\,.
\end{equation}
If we had performed fit $A$ (no $\asq$ term) with the same fit range,
we had obtained
\begin{equation*}
\frac{\langle \bar \psi \psi \rangle^{\msbar}}{N_f}(2 \mathrm{~GeV}) 
= -(335 ~ (37) \mathrm{~MeV})^3 \,, \hspace{1.0 cm} 
m_q^{\msbar}(2 \mathrm{~GeV}) = 3.0 ~ (4) \mathrm{~MeV}\,.
\end{equation*}
Thus, the difference is well covered by the
purely statistical error which is quoted here. 
It is worthwhile to note that in both cases we were able to get
reasonable stable fits of the fitted parameters $\pbp$ and $m_q$ when
including the
$\asq$ term in the fit with fixed values. We conclude
that although we were not able to fit both
nonperturbative condensates independently we get consistent results for $\pbp$ and
$m_q$ with and without the contribution of $\asq$. 


\subsection{Fit $C$:  Determination of the Quark and Gluon Condensates.}
\label{sec:fit_C}

A drawback of the previous fitting strategies is that the sensitivity 
of the scalar to the vector form factor ratio, Eq.~(\ref{eq:ratio}), 
to the value of the gluon condensate is quite limited. As a matter of fact 
the gluon condensate is either assumed to be zero (as in the fits $A$ and $B$) 
or fixed at the value extracted from the analysis of the data for the form 
factor $\hat{V}$ extrapolated to the chiral limit (as in the fits $A'$ and $B'$). 
The main reason for such a limited sensitivity is that, by expanding the 
denominator in Eq.~(\ref{eq:ratio}), the power corrections depending 
on $\asq$ appear always multiplied by the quark mass $m_q$, which is a 
small quantity. Moreover, the results of the previous fits suggest that 
the values of quark and gluon condensates are anti-correlated to each other.
Therefore, in fit $C$ we use the data for the corrected form factors 
$\hat{S}$ and $\hat{V}$, 
separately, in order to determine simultaneously both condensates. 
The price to be paid is the introduction of the renormalization constant 
of the quark field, $Z_q$, which is treated as a free parameter for each 
value of the lattice spacing.

At the same time one of the main outcome of the previous analyses is that 
the discretization effects on the quark mass and on the condensates appear 
to be proportional to the square of the lattice spacing and, moreover, the 
light $u/d$ quark mass obtained in the continuum limit and at the physical 
point turns out to be in good agreement with the existing estimates made by 
the ETMC (see Ref.~\cite {Baron:2009wt}). Thus, we try also
an alternative description of the lattice artifacts by fixing 
the physical quark masses $m_q$ to their $Z_P$-based values and by performing 
in Eqs.~(\ref{opescalar}) and (\ref{opevector}) the following replacements:
\begin{equation}
\begin{split}
      m_q & \to \frac{1}{Z_P ~ a} (a \mu_q) \cdot 
          \left(1 + D_m \frac{a^2}{r_0^2} \right)\,, \nonumber \\
      \pbp & \to  \pbp \cdot 
       \left(1 + D_{\bar{\psi} \psi} \frac{a^2}{r_0^2} \right)\,, 
      \nonumber \\
      \asq & \to  \asq \cdot \left(1 + D_{A^2} \frac{a^2}{r_0^2} \right)\,, 
      \nonumber 
\end{split}
\end{equation}
where $D_m$, $D_{\psi \bar{\psi}}$ and $D_{A^2}$ are free parameters 
(independent on the lattice spacing).
Finally, two discretization terms of the form $D_{S(V)} a^2 k^2$, where 
$D_{S(V)}$ is a free parameter, are added to Eqs.~(\ref{opescalar}) and 
(\ref{opevector}), respectively, to take into account possible discretization 
effects proportional to the squared momentum. The impact of such terms on the 
extraction of the quark and gluon condensates turns out to be quite limited.

A total of $640$ data points are analysed using $11$ free parameters, obtaining a 
$\chi^2$ per degree of freedom of $\sim 0.9$. The quark and gluon 
condensates turn out to be:
\begin{equation}
\begin{split}
      \frac{\pbp^{\msbar}}{N_f} & =  
              -( 270 ~ (15) ~ (20) \mathrm{~MeV})^3 \,, \nonumber \\
      \asq^{\msbar} & = 0.56 ~ (06) ~ (12) \mathrm{~GeV}^2\,, \nonumber 
\end{split}
\end{equation} 
where the second error is the systematic one reflecting the uncertainty 
in the truncation of the perturbative series.
The value obtained for $\asq$ agrees with the one used in fits $A^{\prime}$ 
and $B^{\prime}$. Furthermore, our estimates for $\asq$ agree with the result 
$g^2 \langle A^2\rangle_{\mu=10\, \mathrm{GeV}}^{\msbar}=
2.01~(11)~(^{+0.61}_{-0.73}) \mathrm{~GeV}^2 $ obtained in 
Ref.~\cite{Blossier:2010vt} from the analysis of the Landau gauge quark 
propagator as in the present paper. The latter, once evoluted at the scale 
$\mu = 2 \mathrm{~GeV}$, corresponds to 
$\langle A^2\rangle_{\mu=2 \, {\mathrm GeV}}^{\msbar}=
0.67~(04)~(^{+0.20}_{-0.24}) \mathrm{~GeV}^2 $.

The result for the quark condensate as obtained from fit $C$ is a bit below 
the results obtained in the previous fits. This means that different treatments 
of the quark mass dependence of the leading term of the OPE of the quark 
propagator may lead to a systematic effect of $\sim 30$ MeV on the extracted 
value of the quark condensate.

Overall we conclude from the comparison of primed with non-primed fits
that the contamination of nonperturbative effects in our data 
is inducing errors that are well covered by the uncertainties we
quote.


\begin{table}[t]
  \centering
  \begin{tabular*}{.7\textwidth}{@{\extracolsep{\fill}}c c c}
    quantity & final value & fitting method \\
    \hline\hline
    $\Bigr.\Bigl.\pbp/N_f$ &  $-(335 ~  (37) ~  (35) \mathrm{~MeV})^3 $ & \ Fit $A$ \\
    $\Bigr.\Bigl.m_q$ &  $\bf 3.0 ~  (4) ~ (2) \mathrm{\bf ~MeV} $ &   \\
    \hline
    $\Bigr.\Bigl.\pbp/N_f$ &  $\bf - (299 ~ (26) ~ (29) \mathrm{\bf ~MeV})^3$  & \ Fit $B$ \\
    \hline
    $\Bigr.\Bigl.\pbp/N_f$ &  $-(324  ~ (37) ~ (34) \mathrm{~MeV})^3$ &  \\
    $\Bigr.\Bigl.m_q$ &  $ 3.0  ~ (4) ~ (2) \mathrm{~MeV}$ &  \ Fit $A'$ \\
    $\Bigr.\Bigl.\asq$ &  $(0.65 ~ (09) ~ (17) \mathrm{~GeV})^2$  & \\
    \hline
    $\Bigr.\Bigl.\pbp/N_f$ &  $- (294 ~ (25) ~ (28) \mathrm{~MeV})^3$  & \ Fit $B'$ \\
    $\Bigr.\Bigl.\asq$ &  $(0.65 ~ (09) ~ (17) \mathrm{~GeV})^2$  &  \\
    \hline
    $\Bigr.\Bigl.\pbp/N_f$ &  $-(270  ~ (15) ~ (20) \mathrm{~MeV})^3 $ & \ Fit $C$ \\
    $\Bigr.\Bigl.\asq$ &  $(0.56  ~ (06) ~ (12) \mathrm{~GeV})^2 $ &  \\
    \hline\hline
  \end{tabular*}
  \caption{
Final values of our fit parameters $\pbp$, $m_q$ and $\asq$, all 
in the $\msbar$ scheme at scale $\mu = 2 \mathrm{~GeV}$. 
The value $m_q$ corresponds to the physical point, whereas $\pbp$
and $\asq$ are understood to be defined in the chiral limit. We emphasize  
that (only) the fit strategies $B$ and $B^{\prime}$ provide an explicite
extrapolation of the data to the chiral limit. This is why the $B$-fit 
value for $\pbp$ is quoted in Eq.~(\ref{eq:final_condensate}).
}
  \label{tab_finalresults}
\end{table}


\section{Conclusions}
\label{sec:conclude}

We have presented a study of the quark propagator on the lattice using
$N_f=2$ Wilson twisted mass fermions. By comparing the numerical data
to perturbative series we were able to determine estimates for
the quark mass and the chiral condensate, which both are fundamental
parameters of QCD. 
A summary of the results for the different analysis methods can be found 
in Table~\ref{tab_finalresults}. From these results, we obtain our final 
estimates printed in bold font in Table~\ref{tab_finalresults} and
quoted in Eqs.~(\ref{eq:final_mass}) and (\ref{eq:final_condensate})  
and in the abstract. 
The results we obtain are well compatible with other lattice determinations 
obtained using alternative approaches, while the errors we quote
appear to be larger. 
However, we believe that these errors give a fair estimate in particular 
of the systematic uncertainties involved in the kind of analysis we
applied in this paper.

Therefore, we conclude that a combined perturbative and lattice analysis of the
quark propagator is possible with recent lattice data, even if the
errors are still large. Smaller values of the lattice spacing are
desirable, since they would allow us to include larger values of the
momenta in the analysis.

We have also studied nonperturbative contaminations of our results at
small values of the momenta, which is in the literature often
interpreted as a contribution of the gluon condensate $A^2$. We do see
contributions from such terms, which are, however, not stable over the 
order in perturbation theory. In fact, in the continuum limit the
contribution decreases with increasing order in $\alpha_s$, and we do
not observe any saturation as is visible in Fig.~\ref{fig:10}. 
Still, our value for the gluon condensate is compatible with the
findings reported in Ref.~\cite{Blossier:2010vt}.

\section*{Acknowledgements}

We would like to thank 
Konstantin G. Chetyrkin, Johann H. K\"uhn, and Karl Jansen for initiating 
this work and for useful discussions. We thank all members of the ETM 
Collaboration for the most fruitful collaboration. We thank Martha Constantinou 
for communicating her one-loop lattice perturbation theory results prior to 
publication and Benoit Blossier for discussions.
This work has been supported in part by the DFG Corroborative Research 
Center SFB/TR9 as well as by the DFG and the NSFC through funds provided to 
the sino-german CRC 110. F.B. acknowledges financial support by the DFG-funded 
Graduate School GK 1504.   
V.L. and S.S. thank MIUR (Italy) for partial support under the contract PRIN08.
For data generation we have used the open source tmLQCD software
suite~\cite{Jansen:2009xp} and bQCD.
In our analysis we have been relying on the open source statistics 
package R~\cite{Rcite}.

\bibliographystyle{h-physrev5}
\bibliography{literatur}

\end{document}